\documentclass{nature}

\usepackage{graphicx}
\usepackage{bm}
\usepackage{color}
\usepackage{ulem}

\title{Experimental observation of bulk nodal lines and electronic surface states in ZrB$\bm{_2}$}

\author{Rui Lou$^{1,*}$, Pengjie Guo$^{1,*}$, Man Li$^{1,2,}${\footnote{These authors contributed equally to this work.}} ,
Qi Wang$^1$, Zhonghao Liu$^3$, Shanshan Sun$^1$, Chenghe Li$^1$, Xuchuan Wu$^1$, Zilu Wang$^1$, Zhe Sun$^4$, Dawei Shen$^3$,
Yaobo Huang$^{2,}${\footnote{huangyaobo@sinap.ac.cn}} , Kai Liu$^{1,}${\footnote{kliu@ruc.edu.cn}} , Zhong-Yi Lu$^1$,
Hechang Lei$^{1,}${\footnote{hlei@ruc.edu.cn}} , Hong Ding$^{5,6,7}$ \& Shancai Wang$^{1,}${\footnote{scw@ruc.edu.cn}}}

\begin{document}

\maketitle

\begin{affiliations}
 \item Department of Physics and Beijing Key Laboratory of Opto-electronic Functional Materials $\textsl{\&}$ Micro-nano Devices, Renmin University of China, Beijing 100872, China
 \item Shanghai Synchrotron Radiation Facility, Shanghai Institute of Applied Physics, Chinese Academy of Sciences, Shanghai 201204, China
 \item State Key Laboratory of Functional Materials for Informatics and Center for Excellence in Superconducting Eletronics, SIMIT, Chinese Academy of Sciences, Shanghai 200050, China
 \item National Synchrotron Radiation Laboratory, University of Science and Technology of China, Hefei 230029, China
 \item Beijing National Laboratory for Condensed Matter Physics, and Institute of Physics, Chinese Academy of Sciences, Beijing 100190, China
 \item School of Physical Sciences, University of Chinese Academy of Sciences, Beijing 100190, China
 \item Collaborative Innovation Center of Quantum Matter, Beijing, China
\end{affiliations}

\newpage
\begin{abstract}
  Topological nodal-line semimetals are characterized by the line-contact bulk band crossings and the topological surface states. Breaking
  certain protecting symmetry turns this system into a Dirac semimetal or Weyl semimetal that hosts zero-dimensional isolated nodal points. Recent advances
  in band theory predicted a topological nodal-line semimetal state possessing a new type of nodal line in AlB$_2$-type diborides. Here, we report
  an experimental realization of nodal-line fermions and associated surface states near the Fermi energy in ZrB$\bm{_2}$ by angle-resolved photoemission spectroscopy
  combined with first-principles calculations. The Dirac nodal lines in ZrB$\bm{_2}$ wind into two groups of nodal rings, which are linked together along
  the $\bm{\Gamma}$-$\bm{K}$ direction. We further observe a distinct surface state connecting to each nodal line, indicative of the nontrivial topological nature of
  the bulk nodal lines. Therefore, our results provide convincing experimental evidence of the nodal-line semimetal states in ZrB$\bm{_2}$ both in the bulk
  and on the surface, suggesting ZrB$\bm{_2}$ as a remarkable platform for discovering unique phenomena induced by nodal-line fermions.
\end{abstract}

\section*{Introduction}
The realization of novel quantum states of matter with nontrivial topology beyond topological insulators has become a significant objective in current
condensed-matter physics research\cite{Hasan2010,Qi2011,Weng2014}. Very recently, the discovery of topological semimetals has achieved this goal, which
ignites extensive work focusing on the exotic topological properties and their underlying connection with the electronic structure\cite{Weng2016Rev}.
The topological semimetal states host nontrivial bulk band-crossing points in crystal momentum space\cite{Volovik2009,FangZ2003}. Characterized
by the degeneracy, distribution of the band-crossing points in the Brillouin zone (BZ), and the associated topological boundary
states, the topological semimetals can be classified into three categories: Dirac, Weyl, and nodal-line semimetals. In Dirac and Weyl semimetals, the
bulk nodes are discrete in the BZ and their surface projections are connected by surface Fermi arcs\cite{Weng2016Rev}. While in nodal-line semimetals,
the bulk nodes extend along one-dimensional curves and the corresponding surface states are flat in dispersion according to the previous
nodal-line modelings\cite{RyuS2002}, where the band crossings of a nodal line should occur at zero energy with a constraint chiral symmetry. Hence the
flat surface bands are dubbed the drumhead states. However, the chiral symmetry is not exact in a real crystal, resulting in the nodal line does not
generally occur at a constant energy, thus the associated topological surface states are not flat either\cite{FangC2016,LiR2016}.

The Dirac semimetal and Weyl semimetal states have been theoretically predicted and experimentally verified in various families of compounds\cite{WanX2011,
WangZ2012,Weng2015Weyl,Huang2015,Soluyanov2015,Chang2016,LiuZ2014,Borisenko2014,LvPRX2015,XuScience2015,LvNP2015,YangL2015}. Although there have been several
theoretical proposals for the material realization of topological nodal-line semimetal states\cite{Weng2016Rev,FangC2016,Burkov2011,YangS2017,YuR2017}, the
conclusive experimental proof remains absent until recent angle-resolved photoemission spectroscopy (ARPES) measurements on TiB$_2$\cite{TiB2017}, showing a
tangible realization of bulk nodal-line fermions. Since the surface states associated with the nodal lines are not observed in TiB$_2$\cite{TiB2017}, the confirmation
of the topological nodal-line semimetal state by the coexistence of bulk evidence and surface signature is still desired. In this work, we investigate the
electronic structure of ZrB$_2$, which is predicted to host similar nodal-line configurations and surface states to that of TiB$_2$\cite{ZhangX2017,FengX2017}. By using
ARPES and first-principles calculations, we clearly observe two groups of nodal rings embedded in different mirror planes. These rings are further found to be
linked with each other along the $\Gamma$-$K$ direction. More importantly, we identify distinct surface states emanating from the bulk nodal lines, the surface
signatures demonstrate the nontrivial topology of the nodal-line semimetal states in ZrB$_2$.

Compared with the previous nodal-line candidates CaAgAs, PbTaSe$_2$, and the ZrSiS family\cite{WangX2017,BianG2016Pb,Schoop2016,Takane2016,LouZST2016,
Hosen2017}, ZrB$_2$ has the following advantages: (1) the whole band-crossing features forming the nodal lines are clearly resolved below the Fermi
energy ($E_F$) in ZrB$_2$, while the nodes of CaAgAs\cite{WangX2017}, PbTaSe$_2$\cite{BianG2016Pb}, and the ZrSiS family\cite{Schoop2016,Takane2016,
LouZST2016,Hosen2017} cannot be observed by ARPES with the crossings located above $E_F$; (2) the nodal-line fermions and their
connections with the surface states are
observed without any interference from other bands in ZrB$_2$, while in PbTaSe$_2$\cite{BianG2016Pb} and the ZrSiS family\cite{Schoop2016,Takane2016,LouZST2016,
Hosen2017}, the observation of bulk nodal lines is seriously obstructed by surface states. Thus, our experimental discovery in ZrB$_2$ establishes a unique system that has
the conclusive evidence of topological nodal-line semimetal states both in the bulk and on the surface.

\section*{Results}
\subsection{Structure and transport properties of ZrB$\bm{_2}$.}
As illustrated in Fig. 1(a), ZrB$_2$ crystallizes in a hexagonal lattice system with the space group $P$6/$mmm$ (No. 191)\cite{Post1954}. The
corresponding bulk and (001)-projected surface BZs are shown in Fig. 1(b), where the $\Gamma$-$K$-$M$ and $\Gamma$-$A$-$H$ planes are two mirror
planes of $D_{6h}$ group. The flat and shiny surface [inset of Fig. 1(c)] for our ARPES measurements is the (00$l$) plane, demonstrated by the
single crystal x-ray diffraction (XRD) pattern in Fig. 1(c). To further check the chemical composition of our samples, we display the core-level
photoemission spectrum in Fig. 1(d), where the characteristic peaks of Zr-4$p$, Zr-5$s$, and B-2$p$ orbitals are clearly revealed.

The magnetic field dependences of the resistivity ($\rho_{xx}$) and Hall resistivity ($\rho_{xy}$) at $T$ = 2.5 K are depicted in Fig. 1(e) and the
inset of Fig. 1(e), respectively. By fitting the experimental curves with the two-carrier model\cite{XiaB2013}, the estimated densities of electron-
and hole-type carriers are 1.156(5)$\times$10$^{21}$ and 1.153(5)$\times$10$^{21}$ cm$^{-3}$, respectively, comparable to that of TiB$_2$ ($n_e$, $n_h$
$\sim$ 10$^{21}$ cm$^{-3}$)\cite{TiB2017} and the ZrSiS family ($n_e$, $n_h$ $\sim$ 10$^{20}$ cm$^{-3}$)\cite{HuJ2016,HuJ2017}. Figure 1(f) presents an
overview of the calculated bulk band structure for ZrB$_2$ without the spin-orbit coupling (SOC) effect. Four Dirac-like band-crossing features, which are denoted as $\alpha$, $\beta$, $\gamma$,
and $\delta$, respectively, give rise to two groups of nodal rings embedded in different mirror planes. When the SOC is included in the calculation, small energy gaps ($\sim$40 meV) open at these band crossings\cite{WangQ2017}. We will proceed to a detailed discussion on the
nodal-line semimetal states by systematic electronic structure investigations in the following.

\subsection{Fermi surface (FS) topologies in four high-symmetry planes.}
We present the measured and calculated FS topologies in Fig. 2. Figures 2(d)-2(f) show the FSs recorded with three different photon energies, $h\nu$ = 132, 94,
and 70 eV, which are close to the $k_z$ $\sim$ $\pi$ (FS3), $\sim$ 0 (FS1), and $\sim$ $-\pi$ (FS2) planes, respectively. By continuously varying the photon
energy, we are able to map out the FS in the $\Gamma$-$A$-$H$ plane (FS4), as displayed in Fig. 2(b). From the FS1 in the $k_z$ $\sim$ 0 plane [Fig. 2(e)], the
$r_1$ nodal rings formed by $\alpha$ and $\beta$ are clearly observed surrounding $K$ points.
From the FS2 and FS3 in the $k_z$ $\sim$ $\pm\pi$
planes [Figs. 2(d) and 2(f)], triangular FSs are resolved at $H$ points, which are attributed as the surface states and discussed later,
but no sign of the $r_1$ nodal ring is revealed.
Another major contrast between FS1 and FS2/FS3 is the absence of intensity around $\Gamma$ point
for the former, and the presence of a circular FS centered at $A$ point for the latter. This difference could be further illuminated by the FS4 in the
$\Gamma$-$A$-$H$ plane [Fig. 2(b)]. The FS
around $A$ point, which agreeing well with the green-coloured FS sheet in the calculation [Figs. 2(c) and 2(g)], arises from a hole-like band presented in Fig. 4.
Characterized by these
features in experiment, the electronic structure exhibits a prominent three-dimensional (3D) nature, as confirmed by the calculated 3D FSs [Fig. 2(g)] and their projections on
the $\Gamma$-$A$-$H$ plane [Fig. 2(c)]. Since the $r_2$ nodal ring in the $\Gamma$-$A$-$H$ plane formed by $\gamma$ and $\delta$ also pass through the Fermi wave
vectors of $\beta$, thereby the $\beta$ band-crossing feature belongs to both the $r_1$ and $r_2$ nodal rings. Consequently, the 3D FS is a nodal-link structure
composed of these two nodal rings. Due to the $k_z$ broadening effect, which is prominent in ARPES measurements with vacuum ultraviolet light\cite{Takane2016,
Strocov2003,Kumigashira1998}, the ARPES spectra reflects the electronic states integrated over a certain $k_z$ region of bulk BZ. Therefore, the (001)-projected $r_2$ nodal rings around $\Gamma$/$A$ points are clearly resolved in FS1-FS3 [Figs. 2(d)-2(f)]. Furthermore, the nodal-link point ($J$) of the $r_1$ and $r_2$ nodal rings along the $\Gamma$-$K$ direction is unambiguously recognized in FS1 [Fig. 2(e)].
All the above observations are consistent with the calculated bulk FSs shown in Figs. 2(c), 2(g), and 2(h).

In addition, we observe that some FS features in the $k_z$ $\sim$ $\pm \pi$ planes (FS2 and FS3) are absent in the calculated bulk FSs, i.e., the
triangular FSs around $H$ points, the FS sheets between two $H$ points, and the hexagonal FS centered at $A$ point [relatively weak in FS3 due to
the matrix element effect and its presence can be proved in Fig. 2(b)]. As illustrated in the $h\nu$-$k_{\parallel}$ [$k_{\parallel}$ is
oriented along the
$\Gamma$-$K$ ($A$-$H$) direction] ARPES intensity plot in Fig. 2(b), the Fermi crossings of the hexagonal FS do not show noticeable photon-energy dependence
over a wide range (65-140 eV), which is an indication of surface state. This result inspires us to carry out surface state spectrum calculations for a (001)-oriented 20-unit-cell-thick
slab terminated by Zr layers. The calculated surface state FSs presented in Fig. 2(i) well reproduce these three experimental FS topologies, demonstrating their surface state origins.

\subsection{Presence of bulk nodal-line fermions below $\bm{E_F}$.}
Based on the discussion above, the $r_1$ nodal rings can only be observed in the $k_z$ $\sim$ 0 plane. This character establishes an excellent system to
separately investigate the bulk nodal lines and electronic surface states. Now we record the near-$E_F$ ARPES spectra along the $\Gamma$-$K$ and $M$-$K$ directions,
which are indicated as cuts 1 and 2 in Fig. 2(a), respectively, in the $k_z$ $\sim$ 0 plane (FS1) to prove the presence of bulk nodal-line fermions in
ZrB$_2$. The intensity plots and corresponding second derivative plots are shown in Figs. 3(a)-3(d). The electronic structure near $E_F$ is only composed
of the $\alpha$ and $\beta$ band-crossing features forming the $r_1$ nodal rings, except a band dispersing from the crossing point of $\beta$
along the $\Gamma$-$K$ direction, which is relatively weak when crossing $E_F$ due to the matrix element effect and thus not resolved in the FS mapping data
in Fig. 2(e). This band is identified as the surface state and discussed later. We then plot the momentum distribution curves (MDCs) of $\beta$ and $\alpha$ in
Figs. 3(e) and 3(f), respectively. The linear dispersions in a large energy range with the crossing points below $E_F$ are unambiguously recognized. By
comparing with the superimposed bulk band calculations in Figs. 3(b) and 3(d), one can obtain a high consistency between experiment and theory, which
provides further evidence for the experimental realization of bulk nodal-line fermions in ZrB$_2$. As for the hole-like band $\sim$2.5 eV below
$E_F$ at $M$ point [Figs. 3(c) and 3(d)], due to the $k_z$ broadening effect in ARPES, it derives from the projection of the band along the $H$-$L$-$H$
direction with a similar feature in Fig. 4(c), which is well reproduced by the calculation.

\subsection{Electronic surface states associated with the bulk nodal lines.}
Next, from the perspective of surface signature, we demonstrate the nontrivial topology of the nodal-line semimetal states in
ZrB$_2$ by investigating the surface states revealed
in the $k_z$ $\sim$ $-\pi$ plane (FS2). In Figs. 4(a), 4(c), and 4(e), we present the experimental band dispersions with overlapped bulk band calculations
along the $A$-$H$, $L$-$H$, and $A$-$L$ directions, as indicated by cuts 3-5 in Fig. 2(a), respectively. Besides the well-reproduced bulk bands, there are
some extra bands (SS1-SS3) not existing in the bulk calculations. According to the experimental FSs measured in the $k_z$ $\sim$ $\pm\pi$ planes
(FS2 and FS3), we can determine the origins of the crossing-$E_F$ ones among these bands, i.e., the SS1 corresponds to the triangular FS around $H$ point,
and the SS2 forms the hexagonal FS surrounding $A$ point and the FS sheet between two $H$ points, respectively. To further understand the three
extra bands, we perform 20-unit-cell-thick slab model calculations with Zr-terminated layers. The calculated band dispersions along the $\bar{\Gamma}$-$\bar{K}$,
$\bar{M}$-$\bar{K}$, and $\bar{\Gamma}$-$\bar{M}$ directions with spectral weight from the topmost unit cell are plotted in Figs. 4(b), 4(d), and 4(f), respectively,
which can reproduce SS1-SS3 very well and confirm their surface state nature.

\section*{Discussion}
In order to clarify the nontrivial topology of the bulk nodal lines realized in ZrB$_2$, which are protected by the mirror-reflection symmetries and
the combination of spatial-inversion symmetry ($P$) and time-reversal symmetry ($T$), i.e., the $P{\cdot}T$ symmetry, here we illustrate the connection
between the nodal lines and the surface states by combining both the bulk and surface observations. We plot the crossing points of $\beta$ and $\alpha$ as
the red and blue solid circles in
Figs. 4(a) and 4(c), whose positions are extracted from the MDCs in Figs. 3(e) and 3(f), respectively. We can observe that the bulk nodes of $\beta$ and
$\alpha$ exactly locate on the loci of the surface states resolved in the $k_z$ $\sim$ $-\pi$ plane. Along the $A$-$H$ direction [Fig. 4(a)], starting from the
node of $\beta$, the SS2 disperses inwards with respect to $A$ point. This behavior resembles that of the extra band observed
along the $\Gamma$-$K$ direction in Fig. 3, showing its (the extra band around $\Gamma$ point) surface state origin associated with the bulk nodal line.
Along the $L$-$H$ direction [Fig. 4(c)], the SS1 passing through the node of $\alpha$ disperses outwards with respect to $L$ point. The bulk nodes
and the surface state spectra are obtained
from two independent measurements carried out in different high-symmetry planes, therefore the discovery of their well match provides convincing evidence
for the topological nature (the bulk-boundary correspondence\cite{FangC2016,Burkov2011,WengHM2015}) of the nodal-line semimetal states in ZrB$_2$.

In summary, our direct experimental observation by ARPES presents conclusive evidence of the bulk nodal-line fermions in ZrB$_2$ with negligible SOC effect.
Under the protection of the mirror-reflection symmetries and the $P{\cdot}T$ symmetry, the
electronic structure hosts two groups of Dirac nodal rings, which are linked together along the $\Gamma$-$K$ direction. Meanwhile, we clearly resolve electronic
surface state emanating from each nodal line, proving
the nontrivial topology of the bulk nodal lines. With the experimental realization of topological nodal-line semimetal states in ZrB$_2$ powerfully supported by
both the bulk evidence and the surface signature, we establish an ideal material platform for studying the novel physics and exotic properties related to nodal
lines.

\begin{methods}
\subsection{Sample synthesis.}
High-quality single crystals of ZrB$_2$ were grown via the Fe flux method\cite{WangQ2017}. The starting elements of Zr (99.95\%), B (99.99\%) and Fe (99.98\%)
were put into an alumina crucible, with a molar ratio of Zr: B: Fe = 3: 6: 17. The mixture was heated up to 1873 K in a high-purity argon atmosphere and then
slowly cooled down to 1623 K at a rate of 4 K/h. The ZrB$_2$ single crystals were separated from the Fe flux using the hot hydrochloric acid solution.
\subsection{Angle-resolved photoemission spectroscopy measurements.}
ARPES measurements were performed at the Dreamline beamline of the Shanghai Synchrotron Radiation Facility using a Scienta D80 analyzer and at the
beamline 13U of the National Synchrotron Radiation Laboratory equipped with a Scienta R4000 analyzer. The energy and angular resolutions were set to
25 meV and 0.2$^{\circ}$, respectively. Fresh surfaces for ARPES measurements were obtained by cleaving the samples \textit{in situ} along the (001)
plane. All spectra presented in this work were recorded at $T$ = 20 K in a working vacuum better than 5$\times$10$^{-11}$ Torr.
\subsection{Band structure calculations.}
First-principles calculations were carried out with the projector augmented wave method\cite{Blochl1994,Kresse1999} as implemented in the Vienna
\textit{ab initio} Simulation Package\cite{Kresse1993}. The generalized gradient approximation of Perdew-Burke-Ernzerhof formula\cite{Perdew1996}
was adopted for the exchange-correlation functional. The kinetic energy cutoff of the plane-wave basis was set to be 420 eV. A 20$\times$20$\times$20
$k$-point mesh was utilized for the BZ sampling. The bulk FSs were investigated by adopting the maximally localized Wannier function method\cite{Souza2001}.
A slab with thickness of twenty unit cells along the [001] direction was used in our surface state calculations. The slabs were separated by a vacuum layer of 20 \AA,
which was sufficient to avoid the interactions between different slabs. SOC effect\cite{WangQ2017} was not taken into account in all the above
calculations.
\subsection{Data availability.}
All relevant data are available from the corresponding authors upon request.
\end{methods}

\section*{Acknowledgements}
The work was supported by the National Key R\&D Program of China (Grants No. 2016YFA0300504 and No. 2017YFA0302903), the National Natural
Science Foundation of China (Grants No. 11774421, No. 11574394, No. 11774423, No. 11774424, No. 11227902, and No. 11704394), and the Chinese
Academy of Sciences (CAS) (Project No. XDB07000000). R.L., K.L., and H.L. were supported by the Fundamental Research Funds for the Central
Universities, and the Research Funds of Renmin University of China (RUC) (Grants No. 17XNH055, No. 14XNLQ03, No. 15XNLF06, and No. 15XNLQ07).
Y.H. was supported by the CAS Pioneer Hundred Talents Program. Z.L. acknowledges Shanghai Sailing Program (No. 17YF1422900).

\section*{Author contributions}
R.L., H.C.L., Y.B.H., H.D., and S.C.W. conceived the experiments.
R.L. performed ARPES measurements with the assistance of M.L., Z.H.L., X.C.W., Z.L.W., Z.S., and D.W.S.
P.J.G., K.L., and Z.Y.L. performed first-principles calculations.
Q.W., S.S.S., C.H.L., and H.C.L. synthesized the single crystals.
R.L., H.C.L., and S.C.W. analysed the experimental data.
R.L. plotted the figures.
R.L. and S.C.W. wrote the manuscript.

\section*{Additional information}

\subsection{Supplementary information}
accompanies the paper on the \textit{npj Quantum Materials} website.
\subsection{Competing interests:}
The authors declare no competing interests.

\begin{figure}
  \begin{center}
  \includegraphics[width=0.72\columnwidth]{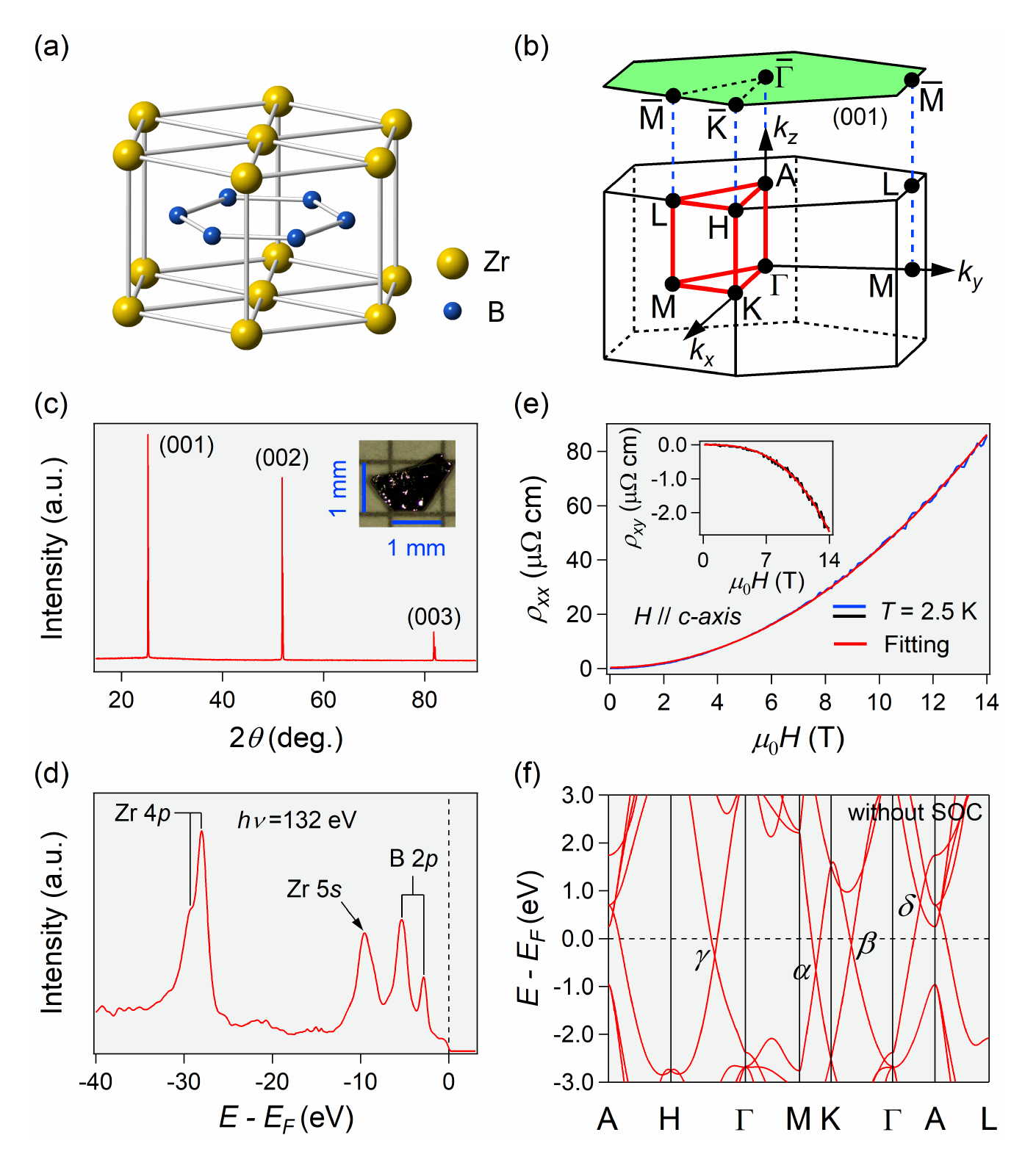}
  \end{center}
  \caption{\textbf{Single crystal and electronic structure of ZrB$\bm{_2}$.}
  (a) Schematic crystal structure of ZrB$_2$.
  (b) Bulk and (001) surface BZs for a hexagonal close-packed structure.
  (c) XRD pattern on the (00$l$) surface. Inset: Picture of a ZrB$_2$ crystal.
  (d) The core-level photoemission spectrum measured with 132-eV photons.
  (e) Magnetic field dependences of the resistivity and Hall resistivity (inset) at $T$ = 2.5 K, displayed as blue and black curves, respectively. The magnetic
      field is parallel to the $c$-axis. The superimposed red curves are the fitting results using the two-carrier model.
  (f) Calculated bulk band structure along the high-symmetry lines without SOC. The near-$E_F$ band-crossing features of $\alpha$, $\beta$, $\gamma$, and $\delta$
      are indicated.}
\end{figure}

\begin{figure}
  \begin{center}
  \includegraphics[width=1\columnwidth]{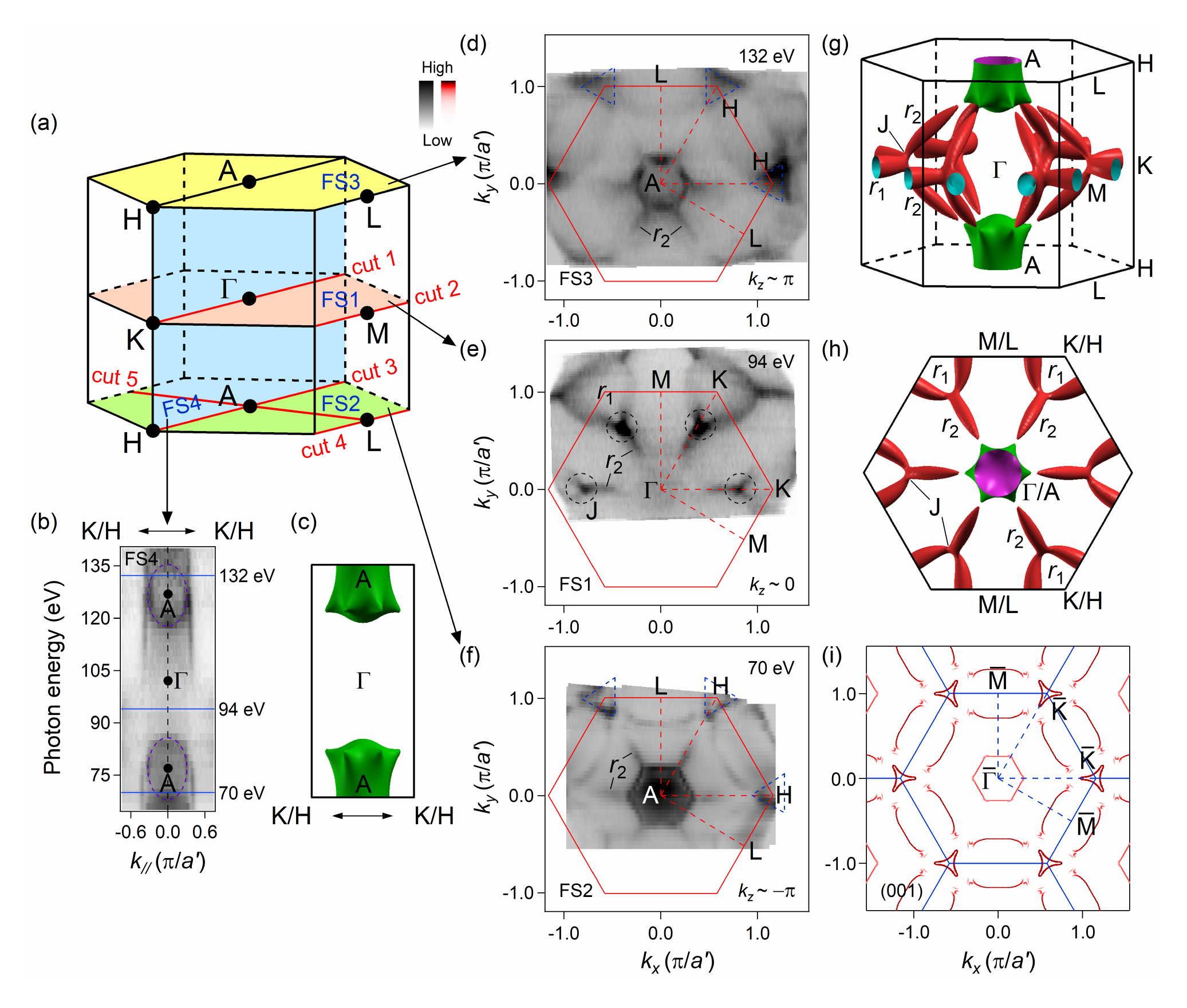}
  \end{center}
  \caption{\textbf{Experimental and theoretical FSs of ZrB$\bm{_2}$.}
  (a) 3D bulk BZ with four high-symmetry planes, which illustrate the momentum locations of the measured FSs in (b) and (d)-(f). Cuts 1, 2 and 3-5 indicate
      the locations of the experimental band structures in Figs. 3 and 4, respectively.
  (b) ARPES intensity map [FS4 in (a)] at $E_F$ in the $h\nu$-$k_{\parallel}$ plane, where $k_{\parallel}$ is oriented along the $\Gamma$-$K$ ($A$-$H$) direction,
      recorded with various photon energies. $a'$ = $\sqrt{3}a$/2 ($a$ = 3.169 \AA). The purple dashed ellipses indicate the hole-like FSs at $A$ points, serving
      as guides to the eye.
  (c) Projections of calculated 3D FSs on the $\Gamma$-$A$-$H$ plane.
  (d)-(f) ARPES intensity plots obtained by $h\nu$ = 132, 94, and 70 eV, showing FS3, FS1, and FS2 indicated in (a), respectively. The red solid lines indicate the
          first BZs projected on the (001) surfaces. Two groups of Dirac nodal rings, $r_1$ and $r_2$ [(001)-projection], are resolved. The
          link points (denoted as $J$ point) of $r_1$ and $r_2$ are indicated by black dashed circles in (e). The blue dashed triangles in (d) and (f) are
          guides to the eye for the triangular FSs around $H$ points.
  (g) Calculated 3D bulk FSs.
  (h) The top view of the calculated 3D FSs, namely the two-dimensional FSs projected on the (001) surface.
  (i) Calculated FS contours of the (001) surface for a 20-unit-cell-thick slab with Zr terminations. The intensity of the red colour scales the spectral weight
      projected to the topmost unit cell. The blue solid lines represent the (001) surface BZs.}
\end{figure}

\begin{figure}
  \begin{center}
  \includegraphics[width=0.76\columnwidth]{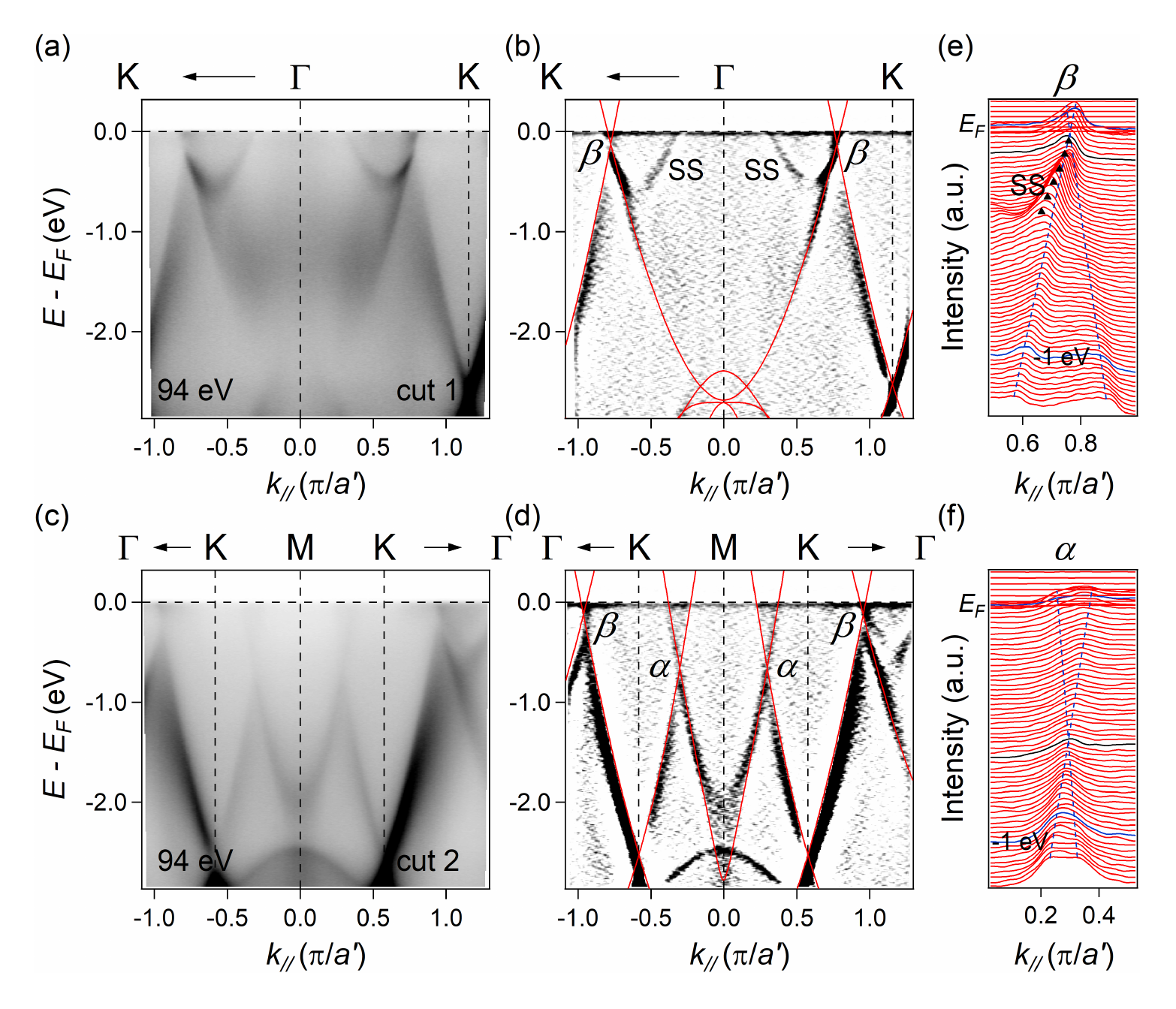}
  \end{center}
  \caption{\textbf{Band structure of the $\bm{r_1}$ nodal ring.}
  (a),(b) Photoemission intensity plot and corresponding second derivative plot along the $\Gamma$-$K$ direction [cut 1 in Fig. 2(a)] recorded with $h\nu$
          = 94 eV, respectively, where SS is short for surface state.
  (c),(d) Same as (a),(b), respectively, but along the $M$-$K$ direction [cut 2 in Fig. 2(a)]. The appended red curves in (b) and (d) are calculated bulk
          band dispersions without SOC. The band-crossing features along the $M$-$K$ and $\Gamma$-$K$ directions are denoted as $\alpha$ and $\beta$, respectively.
  (e),(f) MDC plots of $\beta$ and $\alpha$, which are taken around the right ones in (a) and (c), respectively. The blue dashed lines indicate the linear
          dispersions. The band-crossing points are highlighted by the black curves. The solid triangles in (e) are extracted peak positions, serving as
          guides to the eye for the surface state emanating from the node of $\beta$.}
\end{figure}

\begin{figure}
  \begin{center}
  \includegraphics[width=0.85\columnwidth]{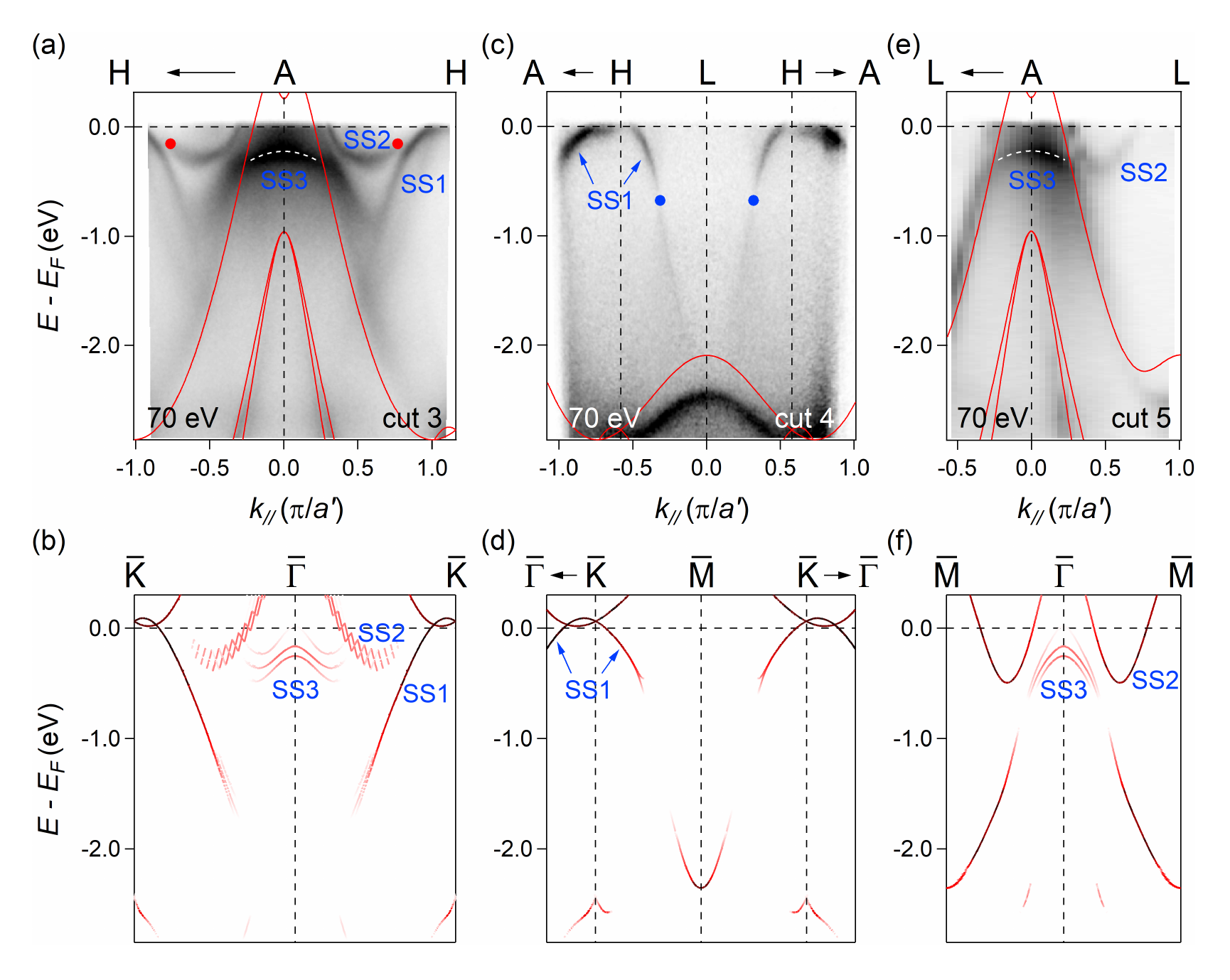}
  \end{center}
  \caption{\textbf{Electronic surface states in the $\bm{k_z}$ $\bm{\sim}$ $\bm{-\pi}$ plane.}
  (a),(c),(e) Photoemission intensity plots along the $A$-$H$, $L$-$H$, and $A$-$L$ directions [cuts 3-5 in Fig. 2(a)] measured with $h\nu$ = 70 eV, respectively.
              The appended red curves are calculated bulk band dispersions without SOC. The white dashed curves in (a) and (e) indicate the SS3 at $A$ point,
              serving as guides to the eye. The red and blue solid circles in (a) and (c) represent the crossing points of $\beta$ and $\alpha$, whose positions
              are determined by the MDCs in Figs. 3(e) and 3(f), respectively.
  (b),(d),(f) Calculated band structure along the $\bar{\Gamma}$-$\bar{K}$, $\bar{M}$-$\bar{K}$, and $\bar{\Gamma}$-$\bar{M}$ directions for a 20-unit-cell-thick
              (001) slab with Zr terminations, respectively. The intensity of the red colour scales the spectral weight projected to the topmost unit cell.}
\end{figure}

\end{document}